\documentclass[12pt,a4paper]{article}
\usepackage{amsmath}
\usepackage{amssymb}
\usepackage{latexsym}
\usepackage{graphicx,color}
\makeatletter
\def\rddots{\mathinner{\mkern1mu\raise\p@%
    \vbox{\kern7\p@\hbox{.}}\mkern2mu%
    \raise4\p@\hbox{.}\mkern2mu\raise7\p@\hbox{.}\mkern1mu}}
\makeatother
\newcommand{\tr}{{\rm tr}}

\begin{document}

\title{\sl Superluminal Group Velocity of Neutrinos : \\
Review, Development and Problems}
\author{
  Kazuyuki FUJII
  \thanks{E-mail address : fujii@yokohama-cu.ac.jp }\\
  International College of Arts and Sciences\footnote{New 
  affiliation after the coming April}\\
  Yokohama City University\\
  Yokohama, 236--0027\\
  Japan
  }
\date{}
\maketitle
\begin{abstract}
  The purpose of this paper is both to provide mathematical reinforcements  
  to the paper [Mecozzi and Bellini : arXiv:1110.1253 [hep-ph]] 
  by taking decoherence into consideration and to present some 
  important problems related.
 
  We claim that neutrinos have superluminality as a latent possibility.
\end{abstract}
%
%
%
%
%     Honbun
%
%
\section{Introduction}
In September 2011 we encountered a remarkable and unbelievable 
paper by the OPERA collaboration \cite{OPERA} that {\bf the speed 
of neutrino exceeds that of light in vacuum}. They measured a collective
speed of mu--neutrino flying from CERN to Gran Sasso Laboratory (
see for example the Fig. 5 in \cite{OPERA}).
However, this result conflicts with special relativity in the most 
basic sense. 

After the paper appeared in the arXiv researchers in the world 
published or are preparing many papers on this topic. 
Some of them agree with the result, while others don't agree 
according to each author's conviction of  the special relativity. 
See the hep-ph in the arXiv. 

We cannot conclude whether this experiment (detection) is correct 
or not at the present time. Of course, it must be checked by other 
experiment teams.

Some researchers protested against the OPERA detection strongly. 
One of main reasons is due to the Kamiokande detection in 1987 
\cite{Kamiokande}. 
It detected lights and neutrinos coming from the Supernova SN1987A 
at almost the same time. If the speed of neutrino is faster than 
that of light in vacuum, it must have detected neutrinos several years 
earlier.

By the way, Mecozzi and Bellini  in \cite{MB}\footnote{K.F gave a small 
contribution to this paper, see Acknowledgments of the paper.}
gave a smart interpretation of the result. They suggested that 
the result is due to {\bf the superluminal group velocity} of neutrinos 
arising from superposition (namely, the neutrino mixing) in 
Quantum Mechanics \cite{PD}. 
The neutrino mixing which is well--known in particle physics is just 
a quantum mechanical phenomenon. 

However, coherence in Quantum Mechanics is affected by 
environments and it could be destroyed in short time. 
A long--distance flight from SN1987A might have destroyed coherence 
of neutrinos, and as a result the superluminal group velocity was lost. 
The paper \cite{MB} offers an interpretation that the OPERA result 
does not conflict with Kamiokande detection.

In this note we provide mathematical reinforcements to the paper \cite{MB} 
in terms of {\bf decoherence} and would like to offer a 
``super--smart" interpretation to the OPERA result \cite{KF}.

\section{Superluminal Group Velocity of Neutrinos}
In this section we review the paper \cite{MB} in detail (because it is 
a bit unclear from the mathematical point of view).

First, we prepare some notation for convenience. Since 
we treat a two level system in the following the target space is  
${\bf C}^{2}=\mbox{Vect}_{{\bf C}}(|\phi_{1}\rangle, |\phi_{2}\rangle)$ 
with bases
\[
|\phi_{1}\rangle=
\left(
\begin{array}{c}
1 \\
0
\end{array}
\right),\quad
|\phi_{2}\rangle=
\left(
\begin{array}{c}
0 \\
1
\end{array}
\right).
\]
Then Pauli matrices $\{\sigma_{x},\ \sigma_{y},\ \sigma_{z}\}$ 
with the identity $1_{2}$
\[
\sigma_{x}=
\left(
\begin{array}{cc}
0 & 1 \\
1 & 0
\end{array}
\right),\quad
\sigma_{y}=
\left(
\begin{array}{cc}
0 & -i \\
i  & 0
\end{array}
\right),\quad
\sigma_{z}=
\left(
\begin{array}{cc}
1 & 0  \\
0 & -1
\end{array}
\right),\quad
1_{2}=
\left(
\begin{array}{cc}
1 & 0 \\
0 & 1
\end{array}
\right)
\]
act on the space.

The three generations of leptons are
\[
\left(
\begin{array}{c}
e         \\
\nu_{e}
\end{array}
\right),\quad
\left(
\begin{array}{c}
\mu        \\
\nu_{\mu}
\end{array}
\right),\quad
\left(
\begin{array}{c}
\tau  \\
\nu_{\tau}
\end{array}
\right).
\]
However, each neutrino is not single but slightly mixed like 
\[
\nu_{\mu}^{\prime}=\cos\Theta\ \nu_{\mu}-\sin\Theta\ \nu_{\tau},\quad
\nu_{\tau}^{\prime}=\sin\Theta\ \nu_{\mu}+\cos\Theta\ \nu_{\tau}.
\]
This $\Theta$ is called the mixing angle in vacuum, which is small enough. 
Therefore real generations are for example
\[
\left(
\begin{array}{c}
\mu                    \\
\nu_{\mu}^{\prime}
\end{array}
\right),\quad
\left(
\begin{array}{c}
\tau                     \\
\nu_{\tau}^{\prime}
\end{array}
\right).
\]
Note that the mixing matrix
\begin{equation}
\label{eq:mixing matrix}
R(\Theta)=
\left(
\begin{array}{cc}
 \cos\Theta & \sin\Theta \\
-\sin\Theta & \cos\Theta
\end{array}
\right)
\end{equation}
is called the Pontecorvo-Maki-Nakagawa-Sakata matrix 
(\cite{Wiki}, \cite{BP}, \cite{MNS}).

Let us start with a model \cite{PM}, \cite{KP}.
Since we treat two neutrinos $\{\nu_{\mu},\ \nu_{\tau}\}$ 
in the paper, the Dirac equation for the two neutrinos can be 
reduced to a Schr{\"o}dinger form written in terms of a two 
component vector of positive energy probability amplitude 
{\bf in the ultra-relativistic limit}.

Then the two neutrino system can be mapped to a two--level 
quantum system with distinct energy eigenvalues along with 
{\bf the assumption of equal fixed momenta} \cite{PM}.

We set that $m_{i}\ (i=1,2)$ are the neutrino masses and $E_{i}$ are 
the energies given by the approximation
\[
E_{i}
=\sqrt{(cp)^{2}+(m_{i}c^{2})^{2}}
\approx cp+\frac{m_{i}^{2}c^{3}}{2p}
=cp+\frac{m_{i}^{2}c^{4}}{2pc}
\]
($p_{1}=p_{2}=p$ from the assumption) and $\Theta$ is the mixing angle. 
The Hamiltonian becomes
\begin{equation}
\label{eq:hamiltonian-1}
H=H(p)
=
\left(cp+\frac{\epsilon_0}{2}\right)1_{2}+
\frac{\epsilon}{2}\{\sin(2\Theta)\sigma_{x}-
\cos(2\Theta)\sigma_{z}\}
\end{equation}
where 
\[
\epsilon_0=\frac{(m_{1}^{2}+m_{2}^{2})c^{4}}{2pc},\quad
\epsilon=\frac{(m_{2}^{2}-m_{1}^{2})c^{4}}{2pc},\quad
E_{0}=cp+\frac{\epsilon_0}{2}.
\]
Here, note that $E_{0}-\frac{\epsilon}{2}=E_{1}$ and 
$E_{0}+\frac{\epsilon}{2}=E_{2}$.

Let us rewrite $H$ in (\ref{eq:hamiltonian-1}) in a familiar 
form by making use of Pauli matrices above
\begin{equation}
\label{eq:hamiltonian-2}
H=
\left(
\begin{array}{cc}
cp+\frac{\epsilon_0}{2}-\frac{\epsilon}{2}\cos(2\Theta) & \frac{\epsilon}{2}\sin(2\Theta) \\
\frac{\epsilon}{2}\sin(2\Theta) & cp+\frac{\epsilon_0}{2}+\frac{\epsilon}{2}\cos(2\Theta) 
\end{array}
\right).
\end{equation}
For this it is easy to obtain the eigenvalues
\begin{equation}
\label{eq:eigenvalues}
E_{\pm}=cp+\frac{\epsilon_0}{2}\pm \frac{\epsilon}{2}
\equiv E_{0}\pm \frac{\epsilon}{2}
\end{equation}
and the corresponding eigenvectors 
\begin{eqnarray*}
|\phi_{-}\rangle
&=&
\left(
\begin{array}{c}
\cos\Theta \\
-\sin\Theta
\end{array}
\right)
=
\cos\Theta|\phi_{1}\rangle-\sin\Theta|\phi_{2}\rangle,  \nonumber \\
|\phi_{+}\rangle
&=&
\left(
\begin{array}{c}
\sin\Theta  \\
\cos\Theta
\end{array}
\right)
=
\sin\Theta|\phi_{1}\rangle+\cos\Theta|\phi_{2}\rangle.
\end{eqnarray*}

If we define a unitary matrix
\[
\left(|\phi_{-}\rangle, |\phi_{+}\rangle \right)
=\left(
\begin{array}{cc}
 \cos\Theta & \sin\Theta \\
-\sin\Theta & \cos\Theta
\end{array}
\right)
=R(\Theta)
\]
then we can make $H$ diagonal like
\begin{equation}
\label{eq:diagonal form}
H
=
R(\Theta)
\left(
\begin{array}{cc}
E_{0}-\frac{\epsilon}{2} & 0  \\
0 & E_{0}+\frac{\epsilon}{2}
\end{array}
\right)
{R(\Theta)}^{\dagger}
=
R(\Theta)
\left(
\begin{array}{cc}
E_{1} & 0  \\
0 & E_{2}
\end{array}
\right)
{R(\Theta)}^{T}
\end{equation}
or in a spectral decomposition form
\begin{equation}
\label{eq:spectral-decomposition}
H=
\left(E_{0}-\frac{\epsilon}{2}\right)|{\phi_{-}}\rangle \langle{\phi_{-}}|+
\left(E_{0}+\frac{\epsilon}{2}\right)|{\phi_{+}}\rangle \langle{\phi_{+}}|.
\end{equation}
These forms are used in the next section.

Next, we label the Hamiltonian by the momentum $p$. Namely,
\begin{equation}
\label{eq:full-hamiltonian}
{\bf H}={\bf H}(p)=H(p)\otimes |p\rangle\langle p|.
\end{equation}
This is in a certain sense the graph of a function.

If we define eigenvectors of ${\bf H}$ in (\ref{eq:full-hamiltonian}) as 
simultaneous ones of both flavor and momentum 
\[
H\otimes {1}|{\phi_{\pm},p}\rangle=E_{\pm}|{\phi_{\pm},p}\rangle, \quad
{1}\otimes \hat{p}|{\phi_{\pm},p}\rangle=p|{\phi_{\pm},p}\rangle,
\]
then we have
\begin{eqnarray}
\label{eq:eigenvectors}
|\phi_{-},p\rangle
&=&
\left(
\begin{array}{c}
\cos\Theta \\
-\sin\Theta
\end{array}
\right)\otimes |p\rangle
=
\cos\Theta|\phi_{1},p\rangle-\sin\Theta|\phi_{2},p\rangle,  \nonumber \\
|\phi_{+},p\rangle
&=&
\left(
\begin{array}{c}
\sin\Theta  \\
\cos\Theta
\end{array}
\right)\otimes |p\rangle
=
\sin\Theta|\phi_{1},p\rangle+\cos\Theta|\phi_{2},p\rangle.
\end{eqnarray}
Therefore, the spectral decomposition of ${\bf H}$ is given by
\begin{equation}
\label{eq:spectral decomposition}
{\bf H}=
\left(E_{0}-\frac{\epsilon}{2}\right)|{\phi_{-},p}\rangle \langle{\phi_{-},p}|+
\left(E_{0}+\frac{\epsilon}{2}\right)|{\phi_{+},p}\rangle \langle{\phi_{+},p}|.
\end{equation}

\vspace{5mm}
Next, we consider a time--evolution of the system. 

\vspace{10mm}
%\begin{center}
\input{time-evolution.fig}
%\end{center}

\vspace{10mm}
The state $|\psi_{t}\rangle$ at time $t$ is given by
\begin{equation}
\label{eq:}
|\psi_{t}\rangle=\int dp^{\prime}e^{-it{\bf H}(p^{\prime})/\hbar}\ |\psi_{0}\rangle
\end{equation}
with the initial state $|\psi_{0}\rangle$, and straightforward calculation 
with (\ref{eq:spectral decomposition}) gives
\begin{equation}
\label{eq:evolution-1}
|\psi_{t}\rangle
=
\int dp^{\prime}e^{-itE_{0}/\hbar}
\left\{
e^{-it\frac{\epsilon}{2\hbar}}\langle{\phi_{+},p^{\prime}}|\psi_{0}\rangle
|{\phi_{+},p^{\prime}}\rangle 
+
e^{it\frac{\epsilon}{2\hbar}}\langle{\phi_{-},p^{\prime}}|\psi_{0}\rangle
|{\phi_{-},p^{\prime}}\rangle 
\right\}
\end{equation}
where $E_{0}=E_{0}(p^{\prime}),\ \epsilon=\epsilon(p^{\prime}),\ 
\epsilon_{0}=\epsilon_{0}(p^{\prime})$ for simplicity.

From now on we assume some conditions :

\noindent
(a)\ \ we start with one neutrino flavor, namely, 
$\langle{\phi_{2},p}|\psi_{0}\rangle=0$,

\noindent
(b)\ \ the initial amplitude of the neutrino waveform is 
$\langle{\phi_{1},p}|\psi_{0}\rangle=\langle{p}|\psi_{0}\rangle$.

\noindent
These assumptions seem to be natural. 

Then it is easy to see that (\ref{eq:evolution-1}) can be rewritten as
\begin{eqnarray}
\label{eq:evolution-2}
&&|\psi_{t}\rangle =
\int dp^{\prime} e^{-itE_{0}/\hbar}\langle{p^{\prime}}|\psi_{0}\rangle \times
\nonumber \\
&&
\left[
e^{-it\frac{\epsilon}{2\hbar}}
\left\{
\sin^{2}\Theta |{\phi_{1},p^{\prime}}\rangle +
\sin\Theta \cos\Theta |{\phi_{2},p^{\prime}}\rangle
\right\}
+
e^{it\frac{\epsilon}{2\hbar}}
\left\{
\cos^{2}\Theta |{\phi_{1},p^{\prime}}\rangle -
\sin\Theta \cos\Theta |{\phi_{2},p^{\prime}}\rangle
\right\}
\right]  \nonumber \\
&{}&
\end{eqnarray}
by use of (\ref{eq:eigenvectors}). 

Now let us start detection of neutrino :

\noindent
(a)\ \ first, we perform a flavor measurement (for example, 
flavor 1),

\noindent
(b)\ \ immediately after (a), we perform a position measurement.

When flavor 1 is detected the collapsed state becomes
\begin{equation}
\label{eq:collapsed state}
|\psi_{t}\rangle \longrightarrow 
|\psi_{t}^{\prime}\rangle
=\frac{1}{\sqrt{{\cal D}}}
\int dp^{\prime}|{\phi_{1},p^{\prime}}\rangle \langle{\phi_{1},p^{\prime}}|\psi_{t}\rangle
=\frac{1}{\sqrt{{\cal D}}}{\cal P}|\psi_{t}\rangle
\end{equation}
where ${\cal D}$ is the normalization factor given by
\begin{equation}
\label{eq:normalization factor}
{\cal D}
=\int dp^{\prime}
\langle{\psi_{t}}|{\phi_{1},p^{\prime}}\rangle \langle{\phi_{1},p^{\prime}}|\psi_{t}\rangle
=\langle\psi_{t}|{\cal P}|\psi_{t}\rangle
\end{equation}
($\langle{\psi_{t}^{\prime}}|\psi_{t}^{\prime}\rangle=1$) and 
\[
{\cal P} \equiv 
\int dp^{\prime}|{\phi_{1},p^{\prime}}\rangle \langle{\phi_{1},p^{\prime}}|
\]
is the projection operator to the flavor 1 state 
(: ${\cal P}^{2}={\cal P},\ {\cal P}^{\dagger}={\cal P}$).

It is in general difficult to perform a position measurement immediately 
after flavor 1 is detected, so we average the positions of neutrinos. 
We believe that this replacement is not so bad.

The expectation value of position measurement on the collapsed state 
$|\psi_{t}^{\prime}\rangle$ is given by
\[
\langle{x_{t}}\rangle=
\int dp^{\prime}\langle{\psi_{t}^{\prime}}|p^{\prime}\rangle
\left(i\hbar \frac{\partial}{\partial p^{\prime}}\right)
\langle p^{\prime}|\psi_{t}^{\prime}\rangle.
\]
This is a kind of definition. Note that
\[
i\hbar \frac{\partial}{\partial p}
\]
is a position operator, because
\[
i\hbar \frac{\partial}{\partial p} \langle{p}|{x}\rangle
=
i\hbar \frac{\partial}{\partial p} e^{-ixp/\hbar}
=xe^{-ixp/\hbar}=x\langle{p}|{x}\rangle.
\]

We must evaluate the expectation value of position 
$\langle{x_{t}}\rangle$. From (\ref{eq:collapsed state}) 
and noting the formula
\begin{equation}
\label{eq:delta function}
\langle{\phi_{1},p^{\prime}}|{\phi_{1},p^{\prime\prime}}\rangle=
\delta(p^{\prime}-p^{\prime\prime})
\end{equation}
we have and set
\begin{equation}
\label{eq:position}
\langle{x_{t}}\rangle=\frac{1}{{\cal D}}
\int dp^{\prime}\langle{\psi_{t}}|{\phi_{1},p^{\prime}}\rangle
\left(i\hbar \frac{\partial}{\partial p^{\prime}}\right)
\langle{\phi_{1},p^{\prime}}|{\psi_{t}}\rangle
\equiv \frac{{\cal N}}{{\cal D}}
\end{equation}
for simplicity.

First, let us calculate $\langle{\phi_{1},p^{\prime}}|{\psi_{t}}\rangle$. 
From (\ref{eq:evolution-2}) and the formula (\ref{eq:delta function}) 
it is easy to see
\begin{equation}
\label{eq:relation}
\langle{\phi_{1},p^{\prime}}|{\psi_{t}}\rangle=
\langle{p^{\prime}}|\psi_{0}\rangle F(p^{\prime})
\end{equation}
where
\begin{equation}
\label{eq:F formula}
F(p^{\prime})=e^{-itE_{0}/\hbar}
\left(
\sin^{2}\Theta e^{-it\epsilon/2\hbar}+\cos^{2}\Theta e^{it\epsilon/2\hbar}
\right).
\end{equation}
Then the normalization factor ${\cal D}$ in (\ref{eq:normalization factor}) 
becomes
\[
{\cal D}=
\int dp^{\prime} |\langle{p^{\prime}}|\psi_{0}\rangle |^{2} |F(p^{\prime})|^{2}.
\]

Now, we make another assumption. The initial distribution of the neutrino 
momentum $|\langle{p^{\prime}}|\psi_{0}\rangle |^{2}$ is narrow with respect 
to $F(p^{\prime})$, and centered on $p^{\prime}=p$ ($p$ is fixed). Namely,
\begin{equation}
\label{eq:initial distribution}
|\langle{p^{\prime}}|\psi_{0}\rangle |^{2} \approx \delta(p^{\prime}-p).
\end{equation}
We believe this one natural. Then
\begin{equation}
\label{eq:value of D}
{\cal D}\approx |F(p)|^{2}.
\end{equation}

For later convenience we calculate (\ref{eq:value of D}). 
Here is an elementary formula

\vspace{3mm}\noindent
{\bf Formula}\ \ for $\alpha, \ \beta \in {\bf R}$
\[
|\alpha e^{-i\theta}+\beta e^{i\theta}|^{2}
=
(\alpha + \beta)^{2}-4\alpha \beta \sin^{2}\theta.
\]
This gives
\begin{eqnarray}
\label{eq:|F(p)|^{2}}
|F(p)|^{2}
&=&
|\sin^{2}\Theta e^{-it\epsilon/2\hbar}+\cos^{2}\Theta e^{it\epsilon/2\hbar}|^{2} 
\nonumber \\
&=&\left(\sin^{2}\Theta +\cos^{2}\Theta\right)^{2}-
4\sin^{2}\Theta\cos^{2}\Theta \sin^{2}\left(\frac{t\epsilon}{2\hbar}\right) 
\nonumber \\
&=&1-\left(2\sin\Theta\cos\Theta\right)^{2}
\sin^{2}\left(\frac{t\epsilon}{2\hbar}\right)    \nonumber \\
&=&1-\sin^{2}(2\Theta)\sin^{2}\left(\frac{t\epsilon}{2\hbar}\right)
\end{eqnarray}
by use of  (\ref{eq:F formula}) (note that $\epsilon=\epsilon(p)$). 

By inserting the equation (\ref{eq:relation}) into ${\cal N}$ in 
(\ref{eq:position}) we have and set
\begin{eqnarray*}
{\cal N}
&=&
\int dp^{\prime} \langle{\psi_{t}}|{\phi_{1},p^{\prime}}\rangle
\left(i\hbar \frac{\partial}{\partial p^{\prime}}\right)
\langle{\phi_{1},p^{\prime}}|{\psi_{t}}\rangle  \\
&=&
\int dp^{\prime} \bar{F}(p^{\prime})\langle{\psi_{0}|p^{\prime}}\rangle
\left(i\hbar \frac{\partial}{\partial p^{\prime}}\right)
\{\langle{p^{\prime}}|\psi_{0}\rangle F(p^{\prime})\}  \\
&=&
\int dp^{\prime} |F(p^{\prime})|^{2}\langle{\psi_{0}|p^{\prime}}\rangle
\left(i\hbar \frac{\partial}{\partial p^{\prime}}\right)
\langle{p^{\prime}}|\psi_{0}\rangle
+
\int dp^{\prime} |\langle{p^{\prime}}|\psi_{0}\rangle |^{2}
\bar{F}(p^{\prime})
\left(i\hbar \frac{\partial}{\partial p^{\prime}}\right)
F(p^{\prime})  \\
&\equiv& {\cal N}_{1}+{\cal N}_{2}
\end{eqnarray*}
for simplicity. Next, let us calculate ${\cal N}_{1}$ and ${\cal N}_{2}$ 
separately.

From
\[
{\cal N}_{1}=\mbox{Re}{\cal N}_{1}+i\mbox{Im}{\cal N}_{1}
\]
we have
\begin{eqnarray}
\mbox{Im}{\cal N}_{1}
&=&
\frac{1}{2i}\left({\cal N}_{1}-\bar{{\cal N}}_{1}\right)  \nonumber \\
&=&
\frac{\hbar}{2}
\int dp^{\prime} |F(p^{\prime})|^{2}
\left\{
\langle{\psi_{0}|p^{\prime}}\rangle
\frac{\partial}{\partial p^{\prime}}
\langle{p^{\prime}}|\psi_{0}\rangle
+
\langle{p^{\prime}}|\psi_{0}\rangle
\frac{\partial}{\partial p^{\prime}}
\langle{\psi_{0}|p^{\prime}}\rangle
\right\}  \nonumber \\
&=&
\frac{\hbar}{2}
\int dp^{\prime} |F(p^{\prime})|^{2}
\frac{\partial}{\partial p^{\prime}}
\left\{
\langle{p^{\prime}}|\psi_{0}\rangle \langle{\psi_{0}|p^{\prime}}\rangle
\right\}  \nonumber \\
&=&
\frac{\hbar}{2}
\int dp^{\prime} |F(p^{\prime})|^{2}
\frac{\partial}{\partial p^{\prime}} |\langle{p^{\prime}}|\psi_{0}\rangle|^{2}
\nonumber \\
&{}& (\mbox{integration by parts})  \nonumber \\
&=&
-\frac{\hbar}{2}
\int dp^{\prime} 
|\langle{p^{\prime}}|\psi_{0}\rangle|^{2}
\frac{\partial}{\partial p^{\prime}} |F(p^{\prime})|^{2} \nonumber \\
&\approx &
-\frac{\hbar}{2}
\int dp^{\prime} \delta(p^{\prime}-p)
\frac{\partial}{\partial p^{\prime}} |F(p^{\prime})|^{2}
=-\frac{\hbar}{2}\frac{\partial}{\partial p} |F(p)|^{2}
\end{eqnarray}
by use of the assumption in (\ref{eq:initial distribution}). Similarly, 
we have
\begin{eqnarray*}
\mbox{Re}{\cal N}_{1}
&=&
\frac{1}{2}\left({\cal N}_{1}+\bar{{\cal N}}_{1}\right)  \\
&=&
\frac{1}{2}
\int dp^{\prime} |F(p^{\prime})|^{2}
\left\{
\langle{\psi_{0}|p^{\prime}}\rangle
\left(i\hbar \frac{\partial}{\partial p^{\prime}}\right)
\langle{p^{\prime}}|\psi_{0}\rangle
-
\langle{p^{\prime}}|\psi_{0}\rangle
\left(i\hbar \frac{\partial}{\partial p^{\prime}}\right)
\langle{\psi_{0}|p^{\prime}}\rangle
\right\}.
\end{eqnarray*}
The range of integration is narrow enough because of 
the assumption in (\ref{eq:initial distribution}), so we 
approximate the integration like

\begin{eqnarray}
\mbox{Re}{\cal N}_{1}
&\approx&
|F(p)|^{2}\ \frac{1}{2}
\int dp^{\prime}
\left\{
\langle{\psi_{0}|p^{\prime}}\rangle
\left(i\hbar \frac{\partial}{\partial p^{\prime}}\right)
\langle{p^{\prime}}|\psi_{0}\rangle
-
\langle{p^{\prime}}|\psi_{0}\rangle
\left(i\hbar \frac{\partial}{\partial p^{\prime}}\right)
\langle{\psi_{0}|p^{\prime}}\rangle
\right\} \nonumber \\
&{}& (\mbox{integration by parts})  \nonumber \\
&=&
|F(p)|^{2}
\int dp^{\prime}
\left\{
\langle{\psi_{0}|p^{\prime}}\rangle
\left(i\hbar \frac{\partial}{\partial p^{\prime}}\right)
\langle{p^{\prime}}|\psi_{0}\rangle
\right\}  \nonumber \\
&=&|F(p)|^{2}\ \langle{x_{0}}\rangle.
\end{eqnarray}
Therefore, we obtain the approximate value
\begin{equation}
\label{eq:N_{1}}
{\cal N}_{1}=\mbox{Re}{\cal N}_{1}+i\mbox{Im}{\cal N}_{1}
\approx 
|F(p)|^{2}\langle{x_{0}}\rangle -\frac{i\hbar}{2}\frac{\partial}{\partial p} |F(p)|^{2}.
\end{equation}

For ${\cal N}_{2}$ we have
\begin{eqnarray}
\label{eq:N_{2}}
{\cal N}_{2}
&=&
\int dp^{\prime} |\langle{p^{\prime}}|\psi_{0}\rangle |^{2}
\bar{F}(p^{\prime})
\left(i\hbar \frac{\partial}{\partial p^{\prime}}\right)
F(p^{\prime})  \nonumber \\
&\approx &
\bar{F}(p)\left(i\hbar \frac{\partial}{\partial p}\right)F(p)
=
i\hbar \bar{F}(p)\frac{\partial}{\partial p}F(p)
\end{eqnarray}
by use of the assumption in (\ref{eq:initial distribution}). 

Then (\ref{eq:N_{1}}) and (\ref{eq:N_{2}}) give
\begin{eqnarray}
\label{eq:N}
{\cal N}={\cal N}_{1}+{\cal N}_{2}
&=&
|F(p)|^{2}\langle{x_{0}}\rangle -\frac{i\hbar}{2}\frac{\partial}{\partial p} |F(p)|^{2}+
i\hbar \bar{F}(p)\frac{\partial}{\partial p}F(p)  \nonumber \\
&=&
|F(p)|^{2}\langle{x_{0}}\rangle+\frac{i\hbar}{2}
\left\{
\bar{F}(p)\frac{\partial}{\partial p}F(p)-\frac{\partial}{\partial p}\bar{F}(p) F(p)
\right\}
\end{eqnarray}
and (\ref{eq:position}) and (\ref{eq:value of D}) give (the approximate value)
\begin{equation}
\label{eq:value of position}
\langle{x_{t}}\rangle=\frac{{\cal N}}{{\cal D}}=\frac{{\cal N}}{|F(p)|^{2}}=
\langle{x_{0}}\rangle +\frac{i\hbar}{2}
\frac{
\bar{F}(p)\frac{\partial}{\partial p}F(p)-\frac{\partial}{\partial p}\bar{F}(p) F(p)
       }{|F(p)|^{2}}.
\end{equation}

Next, let us calculate the right hand side of (\ref{eq:value of position}) 
by use of (\ref{eq:F formula}) :
\[
F(p)=e^{-itE_{0}/\hbar}
\left(
\sin^{2}\Theta e^{-it\epsilon/2\hbar}+\cos^{2}\Theta e^{it\epsilon/2\hbar}
\right).
\]
Noting $E_{0}=E_{0}(p),\ \epsilon=\epsilon(p)$ we have
\[
\frac{\partial}{\partial p}F(p)
=
-i\frac{t}{\hbar}\frac{\partial E_{0}}{\partial p}F(p)
-i\frac{t}{2\hbar}\frac{\partial \epsilon}{\partial p} e^{-itE_{0}/\hbar}
\left(
\sin^{2}\Theta e^{-it\epsilon/2\hbar}-\cos^{2}\Theta e^{it\epsilon/2\hbar}
\right)
\]
and
\begin{eqnarray*}
\bar{F}(p)\frac{\partial}{\partial p}F(p)
&=&
-i\frac{t}{\hbar}\frac{\partial E_{0}}{\partial p}|F(p)|^{2} \\
&{}&
-i\frac{t}{2\hbar}\frac{\partial \epsilon}{\partial p}
\left(
\sin^{2}\Theta e^{-it\epsilon/2\hbar}-\cos^{2}\Theta e^{it\epsilon/2\hbar}
\right)
\left(
\sin^{2}\Theta e^{it\epsilon/2\hbar}+\cos^{2}\Theta e^{-it\epsilon/2\hbar}
\right) \\
&=&
-i\frac{t}{\hbar}\frac{\partial E_{0}}{\partial p}|F(p)|^{2}
-i\frac{t}{2\hbar}\frac{\partial \epsilon}{\partial p}
\left(\sin^{4}\Theta -\cos^{4}\Theta + ** \right)  \\
&=&
-i\frac{t}{\hbar}\frac{\partial E_{0}}{\partial p}|F(p)|^{2}
-i\frac{t}{2\hbar}\frac{\partial \epsilon}{\partial p}
\left(\sin^{2}\Theta -\cos^{2}\Theta + ** \right) 
\end{eqnarray*}
where $**$ is the terms which will be neglected at the final stage. 
This gives
\begin{eqnarray*}
\frac{i\hbar}{2}
\left(
\bar{F}(p)\frac{\partial}{\partial p}F(p)-\mbox{c.c.}
\right)
&=&
t\frac{\partial E_{0}}{\partial p}|F(p)|^{2}+
\frac{t}{2}
\frac{\partial \epsilon}{\partial p}\left(\sin^{2}\Theta -\cos^{2}\Theta\right)  \\
&=&
t\frac{\partial E_{0}}{\partial p}|F(p)|^{2}-
\frac{t}{2}
\frac{\partial \epsilon}{\partial p}\cos(2\Theta)
\end{eqnarray*}
and $\langle{x_{t}}\rangle$ in (\ref{eq:value of position}) is given by
\begin{equation}
\label{eq:}
\langle{x_{t}}\rangle=
\langle{x_{0}}\rangle +t\frac{\partial E_{0}}{\partial p} -
\frac{t}{2}\frac{\cos(2\Theta)}{|F(p)|^{2}}
\frac{\partial \epsilon}{\partial p}.
\end{equation}

\vspace{5mm}\noindent
{\bf Definition}\ \ The group velocity $v_{g}$ is given by
\[
\langle{x_{t}}\rangle - \langle{x_{0}}\rangle =v_{g}t
\Longleftrightarrow v_{g}=
\frac{\langle{x_{t}}\rangle - \langle{x_{0}}\rangle}{t}.
\]
Therefore it becomes
\begin{equation}
\label{eq:group velocity}
v_{g}=\frac{\partial E_{0}}{\partial p}-\frac{1}{2}
\frac{\cos(2\Theta)}{|F(p)|^{2}}\frac{\partial \epsilon}{\partial p}
\end{equation}
from the result above.

Remembering
\[
E_{0}=cp+\frac{\epsilon_0}{2},\quad
\epsilon_0=\frac{(m_{1}^{2}+m_{2}^{2})c^{4}}{2pc},\quad
\epsilon=\frac{(m_{2}^{2}-m_{1}^{2})c^{4}}{2pc}
\]
from (\ref{eq:hamiltonian-1}) simple calculation gives
\[
\frac{\partial E_{0}}{\partial p}=c-\frac{\epsilon_{0}}{2p},\quad
\frac{\partial \epsilon}{\partial p}=-\frac{\epsilon}{p}
\]
and by inserting the above into (\ref{eq:group velocity}) we have
\begin{equation}
v_{g}=c-\frac{\epsilon_{0}}{2p}+{\cal S}\frac{\epsilon}{2p}
\end{equation}
where ${\cal S}$ is given by
\begin{equation}
\label{eq:S}
{\cal S}=\frac{\cos(2\Theta)}{|F(p)|^{2}}=
\frac{\cos(2\Theta)}
{1-\sin^{2}(2\Theta)\sin^{2}\left(\frac{t\epsilon}{2\hbar}\right)}
\end{equation}
from (\ref{eq:|F(p)|^{2}}). 

As a result we obtain

\vspace{5mm}\noindent
{\bf Theorem} (Mecozzi and Bellini)\ \ The group velocity $v_{g}$ 
is given by
\begin{equation}
\label{eq:theorem}
v_{g}=c-\frac{\epsilon_{0}}{2p}+
\frac{\cos(2\Theta)}
{1-\sin^{2}(2\Theta)\sin^{2}\left(\frac{t\epsilon}{2\hbar}\right)}\frac{\epsilon}{2p}.
\end{equation}

\vspace{5mm}
Note that the term ${\cal S}$ was obtained from a quantum effect (the neutrino 
mixing) and this plays a definite role in {\bf Superluminal Group Velocity}.

Let us analyze the theorem. For the purpose we set
\[
\alpha\equiv \sin^{2}\left(\frac{t\epsilon}{2\hbar}\right)
\]
to look for some condition satisfying ${\cal S}>1$. Namely,
\begin{eqnarray*}
&&{\cal S}=\frac{\cos(2\Theta)}{1-\alpha\sin^{2}(2\Theta)}>1 \\
\Longleftrightarrow  
&&\cos(2\Theta)>1-\alpha\sin^{2}(2\Theta)=1-\alpha(1-\cos^{2}(2\Theta)) \\
\Longleftrightarrow
&&(1-\cos^{2}(2\Theta))\alpha>1-\cos(2\Theta)\quad (\mbox{excepting} \ 
1=\cos(2\Theta))  \\
\Longleftrightarrow
&&\alpha>\frac{1}{1+\cos(2\Theta)}>\frac{1}{2}.
\end{eqnarray*}
Therefore, we obtain
\begin{equation}
\label{eq:}
\alpha=\sin^{2}\left(\frac{t\epsilon}{2\hbar}\right)>\frac{1}{2}
\ \Longrightarrow\ {\cal S}>1.
\end{equation}

Here, we assume $m_{2}\gg m_{1}$. Then $\epsilon \approx \epsilon_{0}$ 
from (\ref{eq:hamiltonian-1}) and ${\cal S}>1$ gives
\[
v_{g}\approx c+({\cal S}-1)\frac{\epsilon_{0}}{2p}>c.
\]
As a result

\vspace{5mm}\noindent
{\bf Corollary 1}\ \ Under the conditions $\alpha>\frac{1}{2}$ and 
$m_{2}\gg m_{1}$ we have the superluminal group velocity
\begin{equation}
\label{eq:superlunimal}
v_{g}>c.
\end{equation}

\vspace{3mm}\noindent
{\bf Note}\ \ 
In general, the higher the generation, the heavier corresponding mass. 
Therefore, the assumption $m_{2}\gg m_{1}$ is not unnatural.

\vspace{3mm}
If $\Theta=0$ (no neutrino mixing) then ${\cal S}=1$ from 
(\ref{eq:S}) and we have
\[
v_{g}
\approx c-\frac{\epsilon_{0}}{2p}+\frac{\epsilon_{0}}{2p}
=c
\]
under $m_{2}\gg m_{1}$. As a result

\vspace{5mm}\noindent
{\bf Corollary 2}\ \ If $\Theta=0$ we have the usual velocity
\begin{equation}
\label{eq:usual}
v_{g}=c
\end{equation}
under $\alpha>\frac{1}{2}$ and $m_{2}\gg m_{1}$. 

\vspace{5mm}
We cannot help admitting a mechanism which accelerates 
neutrinos arising from the neutrino oscillation. 
How do we interpret the result ?  What is the relation to 
special relativity ? In the last part of the paper \cite{MB} 
they write :

\vspace{3mm}\noindent
``Of course, this does not mean that the speed of a possible 
signal transmitted with a neutrino wave-packet exceeds 
the speed of light, it is just a property that comes from 
the wave-packet deformation caused by the interference 
of the two possible quantum paths that a neutrino may 
follow before reaching the detector".

\vspace{3mm}
Unfortunately, the author cannot understand what they meant 
and therefore present the following

\vspace{3mm}\noindent
{\bf Problem}\ \ Give a mathematical expression to their claim.

\section{Decoherence of the Neutrino Oscillation}
In this section, for the model in the previous section 
we take {\bf decoherence} into consideration in order to 
make it more realistic. Then we can build a bridge between 
Corollary 1 and Corollary 2. 
For a general introduction to this topic see for example 
\cite{BandP}.

First of all we present the following

\vspace{3mm}\noindent
{\bf Problem}\ \ Is there no problem to apply theory of 
decoherence to neutrinos in a long--distance flight ?

\vspace{3mm}\noindent
Although this problem is very subtle, let us proceed to the 
discussion of decoherence.

Since the two neutrino system can be mapped to the 
two level system we prepare some notation from 
Quantum Optics. For
\[
\sigma_{+}\equiv \frac{1}{2}(\sigma_{x}+i\sigma_{y})=
\left(
\begin{array}{cc}
0 & 1 \\
0 & 0
\end{array}
\right),\quad
\sigma_{-}\equiv \frac{1}{2}(\sigma_{x}-i\sigma_{y})=
\left(
\begin{array}{cc}
0 & 0 \\
1 & 0
\end{array}
\right)
\]
it is easy to see
\[
\sigma_{+}\sigma_{-}=
\left(
\begin{array}{cc}
1 & 0 \\
0 & 0
\end{array}
\right),\quad
\sigma_{-}\sigma_{+}=
\left(
\begin{array}{cc}
0 & 0 \\
0 & 1
\end{array}
\right).
\]

Let us remember
\begin{eqnarray}
\label{eq:diagonal form again}
H
&=&
R(\Theta)
\left(
\begin{array}{cc}
E_{1} & 0  \\
0 & E_{2}
\end{array}
\right)
{R(\Theta)}^{T} \nonumber \\
&=&
\left(
\begin{array}{cc}
E_{1}\cos^{2}\Theta+E_{2}\sin^{2}\Theta & (E_{2}-E_{1})\sin\Theta \cos\Theta \\
(E_{2}-E_{1})\sin\Theta \cos\Theta & E_{1}\sin^{2}\Theta+E_{2}\cos^{2}\Theta
\end{array}
\right)
\end{eqnarray}
from (\ref{eq:diagonal form}) and set
\begin{equation}
\label{eq:diagonal matrix}
H_{0}=
\left(
\begin{array}{cc}
E_{1} & 0  \\
0 & E_{2}
\end{array}
\right).
\end{equation}
Note that $H$ and $H_{0}$ are symmetric matrices ($H=H^{T}$,\ 
$H_{0}=H_{0}^{T}$).

To treat decoherence in a correct manner we must change 
models based on from a pure state to a density matrix. 
The general definition of density matrix $\rho$ is given by both 
$\rho^{\dagger}=\rho$ and $\tr{\rho}=1$, so we can write 
$\rho=\rho(t)$ as
\begin{equation}
\label{eq:density matrix}
\rho=
\left(
\begin{array}{cc}
a         & b  \\
\bar{b} & d
\end{array}
\right)
\quad (a=\bar{a},\ d=\bar{d},\ a+d=1).
\end{equation}

The general form of master equation is well--known to be
\begin{equation}
\label{eq:master equation}
\frac{d}{dt}\rho=-i[H, \rho]+D\rho \quad (\Leftarrow \hbar=1\ 
\mbox{for simplicity})
\end{equation}
where
\[
D\rho=
\mu\left(\sigma_{-}\rho\sigma_{+}-\frac{1}{2}\sigma_{+}\sigma_{-}\rho-\frac{1}{2}\rho\sigma_{+}\sigma_{-}\right)
+
\nu\left(\sigma_{+}\rho\sigma_{-}-\frac{1}{2}\sigma_{-}\sigma_{+}\rho-\frac{1}{2}\rho\sigma_{-}\sigma_{+}\right)
\]
and $\mu>\nu>0$ \footnote{we don't know how to determine 
the precise values of $\mu$ and $\nu$ in our system}. 
We must solve the equation.

If we write $H$ in (\ref{eq:diagonal form again}) as
\begin{equation}
\label{eq:diagonal form again 2}
H=
\left(
\begin{array}{cc}
h & k \\
k & l
\end{array}
\right)
\quad (h,\ k,\ l\ \in {\bf R})
\end{equation}
for simplicity, then the master equation above can be rewritten as
\begin{equation}
\label{eq:master equation 2}
\frac{d}{dt}
\left(
\begin{array}{c}
a \\
b \\
\bar{b} \\
d
\end{array}
\right)
=
\left(
\begin{array}{cccc}
-\mu & ik & -ik & \nu                        \\
ik & i(l-h)-\frac{\mu+\nu}{2} & 0 & -ik   \\
-ik & 0 & -i(l-h)-\frac{\mu+\nu}{2} & ik  \\
\mu & -ik & ik & -\nu 
\end{array}
\right)
\left(
\begin{array}{c}
a \\
b \\
\bar{b} \\
d
\end{array}
\right).
\end{equation}
We leave the derivation to readers.

Note and set
\begin{eqnarray*}
&&\left(
\begin{array}{cccc}
-\mu & ik & -ik & \nu                        \\
ik & i(l-h)-\frac{\mu+\nu}{2} & 0 & -ik   \\
-ik & 0 & -i(l-h)-\frac{\mu+\nu}{2} & ik  \\
\mu & -ik & ik & -\nu 
\end{array}
\right) \\
&=&
\left(
\begin{array}{cccc}
0 & ik & -ik & 0        \\
ik & i(l-h) & 0 & -ik   \\
-ik & 0 & -i(l-h) & ik  \\
0 & -ik & ik & 0
\end{array}
\right)
+
\left(
\begin{array}{cccc}
-\mu & 0 & 0 & \nu                 \\
0 & -\frac{\mu+\nu}{2} & 0 & 0  \\
0 & 0 & -\frac{\mu+\nu}{2} & 0  \\
\mu & 0 & 0 & -\nu 
\end{array}
\right)
\equiv \widehat{H}+\widehat{D}.
\end{eqnarray*}
The general solution of (\ref{eq:master equation 2}) 
is given by
\begin{equation}
\label{eq:general solution}
\left(
\begin{array}{c}
a(t) \\
b(t) \\
\bar{b}(t) \\
d(t)
\end{array}
\right)
=
e^{t\left(\widehat{H}+\widehat{D}\right)}
\left(
\begin{array}{c}
a(0) \\
b(0) \\
\bar{b}(0) \\
d(0)
\end{array}
\right).
\end{equation}
However, it is not easy to calculate the term 
$e^{t\left(\widehat{H}+\widehat{D}\right)}$ exactly, so 
we use a simple approximation
\[
e^{t\left(\widehat{H}+\widehat{D}\right)}=
e^{t\left(\widehat{D}+\widehat{H}\right)}\approx 
e^{t\widehat{D}}e^{t\widehat{H}}.
\]
In general, we must use the Zassenhaus formula, see 
for example \cite{CZ}, \cite{Five}. Therefore, we treat 
the approximate solution
\begin{equation}
\label{eq:general approximate solution}
\left(
\begin{array}{c}
a(t) \\
b(t) \\
\bar{b}(t) \\
d(t)
\end{array}
\right)
\approx 
e^{t\widehat{D}}e^{t\widehat{H}}
\left(
\begin{array}{c}
a(0) \\
b(0) \\
\bar{b}(0) \\
d(0)
\end{array}
\right).
\end{equation}

\vspace{3mm}
First, we calculate $e^{t\widehat{D}}$. For the purpose 
we set
\[
K=
\left(
\begin{array}{cc}
-\mu & \nu \\
\mu & -\nu 
\end{array}
\right)
\]
and calculate $e^{tK}$. The eigenvalues of $K$ are 
$\{0,-(\mu+\nu)\}$ and corresponding eigenvectors (
not normalized) are
\[
0\longleftrightarrow 
\left(
\begin{array}{c}
\nu \\
\mu 
\end{array}
\right),\quad
-(\mu+\nu)\longleftrightarrow 
\left(
\begin{array}{c}
1  \\
-1
\end{array}
\right).
\]
If we define the matrix
\[
O=
\left(
\begin{array}{cc}
\nu & 1   \\
\mu & -1
\end{array}
\right)
\Longrightarrow 
O^{-1}=\frac{1}{\mu+\nu}
\left(
\begin{array}{cc}
1 & 1          \\
\mu & -\nu 
\end{array}
\right)
\]
then it is easy to see
\[
K=
O
\left(
\begin{array}{cc}
0 &                 \\
  & -(\mu+\nu)
\end{array}
\right)
O^{-1}
\]
and
\[
e^{tK}=
O
\left(
\begin{array}{cc}
1 &                         \\
   & e^{-t(\mu+\nu)}
\end{array}
\right)
O^{-1}
=
\frac{1}{\mu+\nu}
\left(
\begin{array}{cc}
\nu+\mu e^{-t(\mu+\nu)} & \nu-\nu e^{-t(\mu+\nu)}  \\
\mu-\mu e^{-t(\mu+\nu)} & \mu+\nu e^{-t(\mu+\nu)}
\end{array}
\right).
\]
Therefore, we have
\begin{equation}
e^{t\widehat{D}}
=
\left(
\begin{array}{cccc}
\frac{\nu+\mu e^{-t(\mu+\nu)}}{\mu+\nu} & 0 & 0 & \frac{\nu-\nu e^{-t(\mu+\nu)}}{\mu+\nu}  \\
0 & e^{-t\frac{\mu+\nu}{2}} & 0 & 0                                                                                  \\
0 & 0 & e^{-t\frac{\mu+\nu}{2}} & 0                                                                                  \\
\frac{\mu-\mu e^{-t(\mu+\nu)}}{\mu+\nu} & 0 & 0 & \frac{\mu+\nu e^{-t(\mu+\nu)}}{\mu+\nu}
\end{array}
\right)
\approx
\frac{1}{\mu+\nu}
\left(
\begin{array}{cccc}
\nu & 0 & 0 & \nu   \\
0 & 0 & 0  & 0        \\
0 & 0 & 0  & 0        \\
\mu & 0 & 0 & \mu 
\end{array}
\right) 
\end{equation}
if $t$ is large enough ($t\gg 1/\nu$).

%\vspace{3mm}
Next, we calculate $e^{t\widehat{H}}$. Since we need 
some properties of tensor product in the following see 
for example \cite{Five}. We can write the equation as
\[
\widehat{H}=-i\left(H\otimes 1_{2}-1_{2}\otimes H\right)
\quad (\Longleftarrow H=H^{T}).
\]
In fact,
\begin{eqnarray*}
\widehat{H}
&=&-i
\left\{
\left(
\begin{array}{cc}
h & k \\
k & l
\end{array}
\right)
\otimes
\left(
\begin{array}{cc}
1 & 0 \\
0 & 1
\end{array}
\right)
-
\left(
\begin{array}{cc}
1 & 0 \\
0 & 1
\end{array}
\right)
\otimes
\left(
\begin{array}{cc}
h & k \\
k & l
\end{array}
\right)
\right\} \\
&=&-i
\left\{
\left(
\begin{array}{cccc}
h & 0 & k & 0 \\
0 & h & 0 & k \\
k & 0 & l  & 0 \\
0 & k & 0 & l
\end{array}
\right)
-
\left(
\begin{array}{cccc}
h & k & 0 & 0 \\
k & l  & 0 & 0 \\
0 & 0 & h & k \\
0 & 0 & k & l 
\end{array}
\right)
\right\}
=-i
\left(
\begin{array}{cccc}
0 & -k & k & 0       \\
-k & -(l-h) & 0 & k \\
k & 0 & l-h  & -k    \\
0 & k & 0-k& 0
\end{array}
\right).
\end{eqnarray*}

It is well--known that
\[
e^{t\widehat{H}}
=
e^{-it\left(H\otimes 1_{2}-1_{2}\otimes H\right)}
=
e^{-it H\otimes 1_{2}}e^{it 1_{2}\otimes H}
=
\left(e^{-itH}\otimes 1_{2}\right)
\left(1_{2}\otimes e^{itH}\right)
=
e^{-itH}\otimes e^{itH}.
\]
Since
\[
H=
R(\Theta)
\left(
\begin{array}{cc}
E_{1} & 0 \\
0 & E_{2}
\end{array}
\right)
R(\Theta)^{T}
\]
we have
\[
e^{-itH}=
R(\Theta)
\left(
\begin{array}{cc}
e^{-itE_{1}} & 0 \\
0 & e^{-itE_{2}}
\end{array}
\right)
R(\Theta)^{T}
\]
and
\begin{eqnarray*}
e^{t\widehat{H}}
&=&
\left\{
R(\Theta)
\left(
\begin{array}{cc}
e^{-itE_{1}} & 0 \\
0 & e^{-itE_{2}}
\end{array}
\right)
R(\Theta)^{T}
\right\}
\otimes
\left\{
R(\Theta)
\left(
\begin{array}{cc}
e^{itE_{1}} & 0 \\
0 & e^{itE_{2}}
\end{array}
\right)
R(\Theta)^{T}
\right\} \\
&=&
\left(R(\Theta)\otimes R(\Theta)\right)
\left\{
\left(
\begin{array}{cc}
e^{-itE_{1}} & 0 \\
0 & e^{-itE_{2}}
\end{array}
\right)
\otimes
\left(
\begin{array}{cc}
e^{itE_{1}} & 0 \\
0 & e^{itE_{2}}
\end{array}
\right)
\right\}
\left(R(\Theta)\otimes R(\Theta)\right)^{T} \\
&=&
\left(R(\Theta)\otimes R(\Theta)\right)
\left(
\begin{array}{cccc}
1 &  &  &                         \\
  & e^{it(E_{2}-E_{1})} &  &    \\
  &   & e^{-it(E_{2}-E_{1})} & \\
  &   &   & 1
\end{array}
\right)
\left(R(\Theta)\otimes R(\Theta)\right)^{T}.
\end{eqnarray*}

Since
\[
R(\Theta)=
\left(
\begin{array}{cc}
 \cos\Theta & \sin\Theta \\
-\sin\Theta & \cos\Theta
\end{array}
\right)
\]
we have
\[
R(\Theta)\otimes R(\Theta)
=
\left(
\begin{array}{cccc}
\cos^{2}\Theta & \cos\Theta\sin\Theta & \cos\Theta\sin\Theta & \sin^{2}\Theta    \\
-\cos\Theta\sin\Theta & \cos^{2}\Theta & -\sin^{2}\Theta & \cos\Theta\sin\Theta \\
-\cos\Theta\sin\Theta & -\sin^{2}\Theta & \cos^{2}\Theta & \cos\Theta\sin\Theta \\
\sin^{2}\Theta & -\cos\Theta\sin\Theta & -\cos\Theta\sin\Theta & \cos^{2}\Theta
\end{array}
\right)
\]
and, by setting $J=e^{it(E_{2}-E_{1})}=e^{it\epsilon}$ for simplicity, 
\begin{eqnarray*}
e^{t\widehat{H}}
&=&
R(\Theta)\otimes R(\Theta)
\left(
\begin{array}{cccc}
1 &  &  &          \\
  & J &  &        \\
  &   & J^{-1} & \\
  &   &   & 1
\end{array}
\right)
\left(R(\Theta)\otimes R(\Theta)\right)^{T} \\
&=&
\left(
\begin{array}{cccc}
\alpha_{11} & \alpha_{12} & \alpha_{13} & \alpha_{14}  \\
* & * & * & *                                                     \\
* & * & * & *                                                     \\
\alpha_{41} & \alpha_{42} & \alpha_{43} & \alpha_{44}
\end{array}
\right)
\end{eqnarray*}
where
\begin{eqnarray*}
\alpha_{11}&=&\cos^{4}\Theta+(J+J^{-1})\cos^{2}\Theta\sin^{2}\Theta+\sin^{4}\Theta, \\
\alpha_{12}&=&-\cos^{3}\Theta\sin\Theta+J\cos^{3}\Theta\sin\Theta
-J^{-1}\cos\Theta\sin^{3}\Theta+\cos\Theta\sin^{3}\Theta, \\
\alpha_{13}&=&-\cos^{3}\Theta\sin\Theta-J\cos\Theta\sin^{3}\Theta
+J^{-1}\cos^{3}\Theta\sin\Theta+\cos\Theta\sin^{3}\Theta, \\
\alpha_{14}&=&\cos^{2}\Theta\sin^{2}\Theta-(J+J^{-1})\cos^{2}\Theta\sin^{2}\Theta
+\cos^{2}\Theta\sin^{2}\Theta
\end{eqnarray*}
and
\begin{eqnarray*}
\alpha_{41}&=&\cos^{2}\Theta\sin^{2}\Theta-(J+J^{-1})\cos^{2}\Theta\sin^{2}\Theta
+\cos^{2}\Theta\sin^{2}\Theta, \\
\alpha_{42}&=&-\cos\Theta\sin^{3}\Theta-J\cos^{3}\Theta\sin\Theta
+J^{-1}\cos\Theta\sin^{3}\Theta+\cos^{3}\Theta\sin\Theta, \\
\alpha_{43}&=&-\cos\Theta\sin^{3}\Theta+J\cos\Theta\sin^{3}\Theta
-J^{-1}\cos^{3}\Theta\sin\Theta+\cos^{3}\Theta\sin\Theta, \\
\alpha_{44}&=&\sin^{4}\Theta+(J+J^{-1})\cos^{2}\Theta\sin^{2}\Theta+\cos^{4}\Theta.
\end{eqnarray*}
Note that $*$'s in the matrix are elements not used in the latter. 
We leave this derivation to readers. 

\noindent
Here, we list very important relations among $\{\alpha\}$
\begin{equation}
\label{eq: important relations}
\alpha_{11}+\alpha_{41}=1,\quad
\alpha_{12}+\alpha_{42}=0,\quad
\alpha_{13}+\alpha_{43}=0,\quad
\alpha_{14}+\alpha_{44}=1.
\end{equation}

Therefore, from (\ref{eq:general approximate solution}) and 
(\ref{eq: important relations}) we obtain
\begin{eqnarray}
\label{eq:Fujii I}
\left(
\begin{array}{c}
a(t) \\
b(t) \\
\bar{b}(t) \\
d(t)
\end{array}
\right)
&\approx& 
\frac{1}{\mu+\nu}
\left(
\begin{array}{cccc}
\nu & 0 & 0 & \nu   \\
0 & 0 & 0  & 0        \\
0 & 0 & 0  & 0        \\
\mu & 0 & 0 & \mu 
\end{array}
\right)
\left(
\begin{array}{cccc}
\alpha_{11} & \alpha_{12} & \alpha_{13} & \alpha_{14}  \\
* & * & * & *                                                     \\
* & * & * & *                                                     \\
\alpha_{41} & \alpha_{42} & \alpha_{43} & \alpha_{44}
\end{array}
\right)
\left(
\begin{array}{c}
a(0) \\
b(0) \\
\bar{b}(0) \\
d(0)
\end{array}
\right) \nonumber \\
&=& 
\frac{1}{\mu+\nu}
\left(
\begin{array}{cccc}
\nu & 0 & 0 & \nu   \\
0 & 0 & 0  & 0        \\
0 & 0 & 0  & 0        \\
\mu & 0 & 0 & \mu 
\end{array}
\right)
\left(
\begin{array}{c}
a(0) \\
b(0) \\
\bar{b}(0) \\
d(0)
\end{array}
\right) 
\end{eqnarray}
for $t\gg 1/\nu$.

\vspace{3mm}
For $H_{0}$ in (\ref{eq:diagonal matrix})
\[
H_{0}=
\left(
\begin{array}{cc}
E_{1} & 0  \\
0 & E_{2}
\end{array}
\right)
\]
we can perform the same process much easily. 
The master equation is
\[
\frac{d}{dt}
\left(
\begin{array}{c}
a \\
b \\
\bar{b} \\
d
\end{array}
\right)
=
\left(
\begin{array}{cccc}
-\mu & 0 & -0 & \nu                       \\
0 & i(l-h)-\frac{\mu+\nu}{2} & 0 & 0   \\
0 & 0 & -i(l-h)-\frac{\mu+\nu}{2} & 0  \\
\mu & 0 & 0 & -\nu 
\end{array}
\right)
\left(
\begin{array}{c}
a \\
b \\
\bar{b} \\
d
\end{array}
\right)
\]
and the (exact) solution is given by
\[
\left(
\begin{array}{c}
a(t) \\
b(t) \\
\bar{b}(t) \\
d(t)
\end{array}
\right)
=
\left(
\begin{array}{cccc}
\frac{\nu+\mu e^{-t(\mu+\nu)}}{\mu+\nu} & 0 & 0 & \frac{\nu-\nu e^{-t(\mu+\nu)}}{\mu+\nu}  \\
0 & e^{it(l-h)}e^{-t\frac{\mu+\nu}{2}} & 0 & 0                                                                      \\
0 & 0 & e^{-it(l-h)}e^{-t\frac{\mu+\nu}{2}} & 0                                                                     \\
\frac{\mu-\mu e^{-t(\mu+\nu)}}{\mu+\nu} & 0 & 0 & \frac{\mu+\nu e^{-t(\mu+\nu)}}{\mu+\nu}
\end{array}
\right)
\left(
\begin{array}{c}
a(0) \\
b(0) \\
\bar{b}(0) \\
d(0)
\end{array}
\right).
\]
When $t\gg 1/\nu$ we obtain
\begin{equation}
\label{eq:Fujii II}
\left(
\begin{array}{c}
a(t) \\
b(t) \\
\bar{b}(t) \\
d(t)
\end{array}
\right)
\approx 
\frac{1}{\mu+\nu}
\left(
\begin{array}{cccc}
\nu & 0 & 0 & \nu   \\
0 & 0 & 0  & 0        \\
0 & 0 & 0  & 0        \\
\mu & 0 & 0 & \mu 
\end{array}
\right)
\left(
\begin{array}{c}
a(0) \\
b(0) \\
\bar{b}(0) \\
d(0)
\end{array}
\right).
\end{equation}

As a result we have

\vspace{5mm}\noindent
{\bf Theorem}\ \ Two systems (master equations) whose 
Hamiltonians are $H$ and $H_{0}$ have the same asymptotic 
behavior (\ref{eq:Fujii I}) and (\ref{eq:Fujii II}) under  
our approximation.

\vspace{5mm}
This theorem implies

\vspace{5mm}\noindent
{\bf Corollary}\ \ The mixing angle $\Theta$ will become 
$0$ if $t$ is large enough.

\vspace{3mm}
From both this corollary and corollary 2 in the preceding section 
we can conclude that the speed of neutrinos (after a long--distance 
flight) is just that of light in vacuum.

By the way, our calculation in this section is based on a simple 
approximation. This is a bit poor, so we present the following

\vspace{3mm}\noindent
{\bf Problem}\ \ Give the explicit (full) calculation.

\vspace{3mm}
As for interesting topics of decoherence (which is essential 
in Quantum Physics) arising from Quantum Optics or Quantum 
Computation see our papers 
\cite{EFS}, \cite{FS2} and \cite{FS-1}, \cite{FS-2}.

\vspace{3mm}
At the end of this section, one comment is in order.  
It seems to the author that Neutrino Physics gets along with 
Quantum Optics or Quantum Computation, see for example \cite{CW}. 
In order to make some (deep) relations clear further studies 
will be required.

\section{Concluding Remarks}
In this paper we re-examined the paper \cite{MB} by Mecozzi and Bellini 
in detail and tried to give mathematical reinforcements to it by taking 
decoherence into consideration. Our conclusion is
\begin{center}
{\bf Neutrinos are latently superluminal.}
\end{center}
We would like to present the following
 
\vspace{3mm}\noindent
{\bf Problem}\ \ Re-check our result from a different point of view. 

\vspace{3mm}
Let us write once more that our argument is based on {\bf group velocity}. 
Therefore, it is a bit unsatisfactory.

Whether the OPERA experiment is correct or not is not concluded  
at the present time and it must be checked by other experiment teams. 
However, such a check will take time. 
Therefore, it is very important for us to state

\vspace{3mm}\noindent
{\bf Problem}\ \ Make some (inside) questions clear 
from a theoretical point of view. 

\vspace{3mm}\noindent
Regarding papers related to this topic see for example \cite{several}.

\vspace{3mm}
The work is a great challenge to not only (young) Physicists 
but also (young) Mathematicians, so we conclude the paper 
by citing famous sentences by late Steve Jobs \footnote{\ Steven 
Paul Jobs (1955--2011)}
\begin{center}
Stay hungry,\ \ Stay foolish
\end{center}
(see the Concluding Remarks in \cite{KF}).

\vspace{5mm}\noindent 
{\it Acknowledgments}\ \ 
We wish to thank Shin'ichi Nojiri and Ryu Sasaki for useful 
suggestions and comments.

%%%%%%%%%%%%%
%References%
%%%%%%%%%%%%%

\end{document}